\documentclass[aps, prd, reprint, longbibliography, nofootinbib,superscriptaddress, floatfix]{revtex4-2}
\pdfoutput=1

\usepackage{amsmath,amssymb,amsfonts}
\usepackage{mathrsfs}
\usepackage{bbm}
\usepackage{slashed}
\usepackage{graphicx}
\usepackage{comment}
\usepackage{verbatim}
\usepackage[T1]{fontenc}
\usepackage[utf8]{inputenc}
\usepackage[colorlinks]{hyperref}
\usepackage{mathbbol}
\usepackage[dvipsnames,table]{xcolor}
\usepackage[normalem]{ulem}
\usepackage{svg}
\usepackage{flushend}
\usepackage{physics}
\usepackage{multirow}
\usepackage{graphicx}
\usepackage{dcolumn}
\usepackage{bm}
\usepackage{tensor}
\usepackage{comment}
\usepackage[utf8]{inputenc} 
\usepackage{amsthm,amsmath,amssymb,hyperref,mathrsfs}
\usepackage{braket,bm,bbm}
\usepackage{cancel}
\PassOptionsToPackage{normalem}{ulem}
\usepackage{ulem} 
\usepackage{physics}
\usepackage{float}
\usepackage[english]{babel}
\usepackage{graphicx}
\hypersetup{
    colorlinks=false,
    pdfborder={0 0 0},
}
\usepackage{comment}
\usepackage[capitalize]{cleveref}

\newcommand{\qu}[1]{``{#1}''} 

\begin{document} 

\title{Gravitational quantum friction and the information-loss problem}

    \author{Francesco Del Porro}
\email[]{francesco.del.porro@nbi.ku.dk}
\affiliation{Center of Gravity, Niels Bohr Institute, Blegdamsvej 17, DK-2100 Copenhagen \O, Denmark}
\affiliation{Niels Bohr International Academy, Niels Bohr Institute, Blegdamsvej 17, DK-2100 Copenhagen \O, Denmark}

\author{Francesco Di Filippo}
\email[]{difilippo@itp.uni-frankfurt.de}
\affiliation{Institut f\"ur Theoretische Physik, Max-von-Laue-Str.1, 60438 Frankfurt, Germany}

\begin{abstract}
\noindent 
We study the gravitational collapse of a shell, modeled as a coherent state of quantum matter. Without adding any physics beyond semiclassical gravity, we show that the universal gravitational coupling between matter and geometry induces an interaction between the shell and the negative-frequency modes present in the $\ket{in}$ vacuum state. We find that this interaction occurs unavoidably immediately after a trapping horizon forms, making it an universal feature of semiclassical gravitational collapse. The resulting transfer of energy from the collapsing matter to outgoing Hawking radiation acts as a \textit{gravitational quantum friction}, opposing the formation of trapped regions while simultaneously imprinting the information carried by the collapsing matter onto the emitted quanta. We discuss the implications of this mechanism for the black hole information-loss problem and speculate about the possible endpoints of collapse, potentially preventing horizon and singularity formation.
\end{abstract}

\maketitle

\section{Introduction}\label{sect:introduction}
Hawking radiation is one of the most fascinating aspects in gravitational physics \cite{Hawking:1975vcx}. 
It shows a non-trivial interplay between quantum field theory and classical general relativity. In its standard derivation, a minimally coupled test field is quantised on a collapsing spacetime, leading to the emission of a thermal flux and to the gradual evaporation of the black hole. The thermal character of the radiation raises conceptual problems regarding the unitary evolution of the theory and lies at the core of the black hole information-loss problem~\cite{Hawking:1975vcx,Page:1993wv,Unruh:2017uaw,Mathur:2009hf,Buoninfante:2021ijy,Buoninfante:2025gqk}. The absence of correlations among emitted quanta, implies that an initial pure state evolves into a mixed one. 

In the last few decades, several solutions to this problem have been put forward; most of which require extra ingredients beyond Hawking's original setup~\cite{Hawking:1993pb,Susskind:1993if,Giddings:1993vj,Mathur:2005zp,Almheiri:2012rt,Almheiri:2019yqk,Almheiri:2019psf}.

In this work, we present a more conservative possibility, which addresses the problem without modifying the assumptions of \cite{Hawking:1975vcx}.

A crucial ingredient that is usually neglected in the original picture is the intrinsic quantum nature of the collapsing matter itself. The collapsing body is typically treated as a classical source whose only role is to generate the background geometry on which the quantum field propagates.

We argue that once the collapsing matter is treated quantum mechanically, the universal gravitational coupling naturally induces an interaction between the infalling particles and the partner modes contained in the $\ket{in}$ vacuum.

Remarkably, this mechanism requires no ingredients beyond those already present in the standard semiclassical description of gravitational collapse. No exotic physics is introduced, nor is an additional matter sector required: the field responsible for the radiation is the same field that constitutes the collapsing matter.

We show that, as soon as a horizon is formed, the presence of negative-normed modes in the $\ket{in}$ state generates a sort of \textit{gravitational quantum friction}, opposing the persistence of the trapped region due to the evaporation. At the same time, the nonlocal correlations of the vacuum provide a channel through which the quantum information carried by the collapsing matter can be coherently imprinted on the outgoing radiation.

We discuss the relevance of this mechanism for the information-loss problem. We also explore, more speculatively, whether its repeated action could qualitatively modify the late stages of gravitational collapse and play a role in the formation of horizons and singularities \cite{Carballo-Rubio:2019fnb,Carballo-Rubio:2025fnc}.

\section{Setup} 
We consider the gravitational collapse of a free-falling thin shell in a coherent state of a large number of non-interacting particles in general relativity. We work in the units $\hbar=c=1$, without fixing the value of $G$. The gravitational sector is described by the usual Eisntein--Hilbert action
\begin{equation}
    S_g=\frac{1}{16\pi G}\int \sqrt{-g}R\,.
\end{equation}

For concreteness, in the matter sector we consider a scalar field charged under a global $SU(2)$ symmetry group. In the fundamental representation, the matter action reads
\begin{equation} \label{eq:S_phi}
    S_\phi= \int \sqrt{-g}  \phi^\dagger (g^{\mu \nu}\nabla_\mu \nabla_\nu + m^2) \phi
\end{equation}
where $\phi=(\phi_+,\phi_-)$ is a doublet. Being in the fundamental representation, the invariance is given by
\begin{equation}
    \phi \to U \phi \qquad U\in SU(2)
\end{equation}
and we choose to label the doublet with the eigenvalues of $T_3=\frac{\sigma_3}{2}$ such as $T_3 \phi_\pm= \pm \frac 12 \phi_\pm$, being $\sigma_3$ the third Pauli matrix. The introduction of a globally-charged scalar field is mainly motivated by having a notion of \qu{qubit} in our system in the simplest way. An extension to cases with different spin or local gauge symmetries -- with the addition of interactions with gauge fields -- would introduce quantitative but not qualitative differences.

The freely falling shell is made by $N$ qubits of energy~$\omega$. Because of the non interacting nature of the particles, the energy of the shell is given by
$M=N\omega$.

We will only need the metric outside (and by continuity on) the shell which by Birkhoff's theorem \cite{Wald:1984rg} is the Schwarzschild metric
\begin{equation}
    ds^2=-F(r)dv^2+2dvdr+r^2d \mathbb{S}_2\,,
\end{equation}
with $F(r)=1-2GM/r$.

Besides the coherent state of the collapsing matter, the field has a non-trivial mode structure given by the usual $\left|in\right>$ vacuum state \cite{Wald:1975kc}. For simplicity, in the following we consider the vacuum state to be the Unruh state. This is justified as we will focus in the region of spacetime close to the outer trapping horizon where the  $\left|in\right>$ state approximates the Unruh state~\cite{Fabbri:2005mw}.

The state is defined as \cite{Wald:1975kc,Jacobson:2003vx}
\begin{equation}\label{eq:U_state}
    \ket{U}\propto\sum _{i=\pm 1}\sum_{n=0}^{+\infty}\frac{e^{-n\pi \omega/\kappa }}{n!}(a_{\omega,i}^{\dagger,{\rm out}}a_{\omega,-i}^{\dagger,{\rm in}})^n\ket{0}\,,
\end{equation}
where $out$ ($in$) indicates the wavepackets peaked on energy $\omega$ ($-\omega$) with support outside (inside) the horizon and $i=\pm$ labels the $SU(2)$ charge. Here $\ket{0}$ represents the Boulware state \cite{Jacobson:2003vx}.

\begin{figure}
    \centering
    \includegraphics[width=0.48\linewidth]{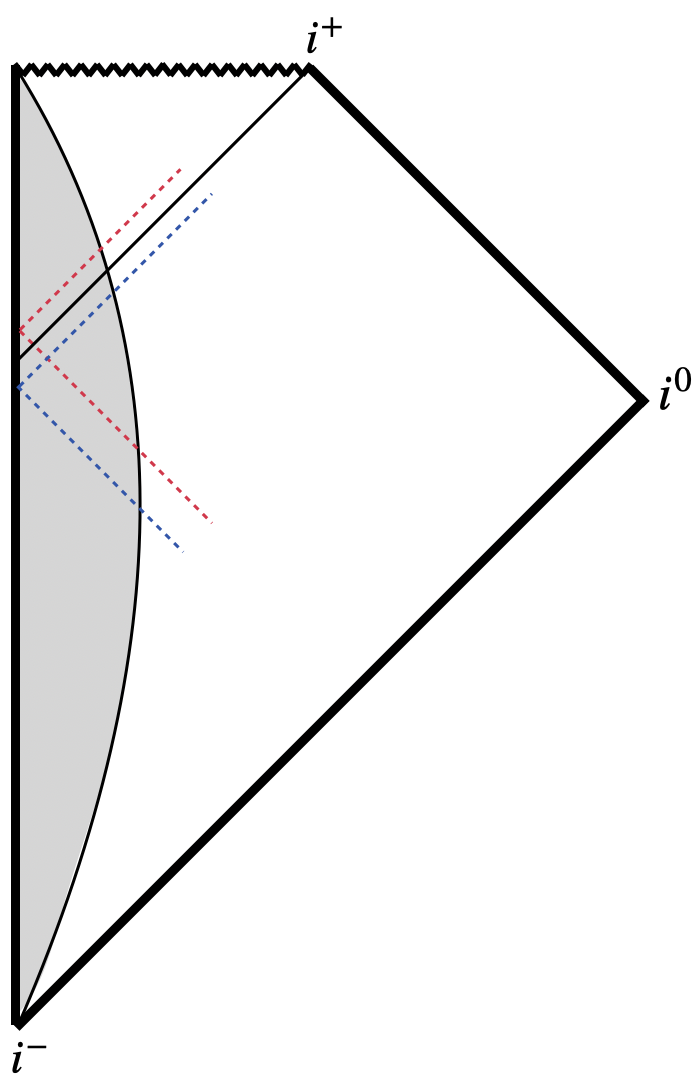}
    \caption{Causal diagram of a black hole spacetime without taking the backreaction into account. Negative-normed Hawking partners (red dashed line) inside the event horizon (black solid line) are entangled with positive-normed Hawking quanta (blue dashed line) outside the horizon. Close to the horizon, the wavepackets travel approximately along outgoing null geodesics.}    \label{fig:no_backreaction}
\end{figure}

Figure~\ref{fig:no_backreaction} shows, in absence of backreaction, the splitting into positive and negative energy modes outside and inside the horizon. For each outgoing wavepacket outside the horizon, the presence of the partner mode with disjoint support is needed for the $\ket{in}$ state to be pure~\cite{Wald:1975kc,Agullo:2024nxg}. 

At the linear level, the coupling with gravity happens through the stress energy tensor
\begin{equation}
    T^{\mu \nu}_\phi:= - \frac{2}{\sqrt{-  g}}\frac{\delta S[ g ;\phi]}{\delta g_{\mu \nu}} \,.
\end{equation}

We take the metric of the infalling shell at mass $M$ and we linearize around it
\begin{equation}
    g_{\mu \nu}= \bar g_{\mu \nu}(M)+  \delta g_{\mu \nu} \,.
\end{equation}
For small perturbation of the geometry for which  $\delta g_{\mu \nu}^2 $ is negligible, the action $S_\phi$ can be expanded at the linear level, leading to
\begin{equation} \label{eq:interaction_L}
    S_\phi[\bar g + \delta g;\phi]=S_\phi[\bar g ;\phi] + \int \delta g_{\mu \nu} T^{\mu \nu}_\phi + \mathcal{O}(\delta g^2)\,.
\end{equation}
For a shell losing a mass $\omega$, the perturbation of the metric reads
\begin{equation}
    \delta g_{\mu \nu} =g_{\mu\nu}(M)-g_{\mu\nu}(M-\omega)=-\frac{2 G\omega}{r} \delta_\mu^v \delta_\nu ^v \,.
\end{equation}
In the matter sector, the mass loss is intended as a scattering between a given qubit $\mathfrak q$ on the shell and the negative-normed mode content of the $\ket{in}$ state. Namely
\begin{equation} \label{eq:matrix_T}
   \bra{\mathcal{S}_{\cancel{\mathfrak{q}}},U_{\cancel{\mathfrak{\bar{q}}}}}\hat T_\phi^{\mu \nu} \ket{\mathcal{S},U}\,,
\end{equation}
where $\mathcal{S}$ represents the shell coherent state including the qubit $\mathfrak{q}$, $\mathcal{S}_{\cancel{\mathfrak{q}}}$ is the shell coherent state without the qubit $\mathfrak{q}$, $U$ is the Unruh state, and $U_{\cancel{\mathfrak{\bar{q}}}}$ is the Unruh state without the mode $\mathfrak{\bar{q}}$. Importantly, the invariance of \cref{eq:S_phi} under $SU(2)$ implies the charge conservation in the matrix element of \cref{eq:matrix_T}. In other words, for a scattering to happen, $\bar{\mathfrak q}$ must have opposite $SU(2)$ charge to $\mathfrak{q}$.

Since the scattering process has support only on the shell's worldtube $X^\mu(\tau)=(v(\tau),r(\tau),\theta,\varphi)$, we model the matrix element \eqref{eq:matrix_T} as
\begin{align} \label{eq:Localized_T}
&\bra{\mathcal{S}_{\cancel{\mathfrak{q}}},U_{\cancel{\mathfrak{\bar{q}}}}}\hat T_\phi^{\mu \nu} \ket{\mathcal{S},U}= \nonumber \\
   &\int r^2(\tau) d \tau d\mathbb{S}_2 \, \frac{\delta^{(4)}(x-X(\tau))}{\sqrt{-g}}  \mathcal T^{\mu \nu}_\phi(x) \,.
\end{align}
The $\delta$-function enforces the thin-shell structure by localizing the integral on the three dimensional surface spanned by the time evolution of the collapsing shell,\footnote{Note that, in our model, the shell acts as a freely falling detector probing the vacuum \cite{Unruh:1976db,Birrell:1982ix}. The main difference with standard treatments is that the gravitationally-mediated coupling introduces derivatives in the interaction terms.} parametrized by its proper time $\tau$
\begin{align} \label{eq:localization}
   & \int \sqrt{-g} d^4x \bra{\mathcal{S}_{\cancel{\mathfrak{q}}},U_{\cancel{\mathfrak{\bar{q}}}}}\hat T_\phi^{\mu \nu} \ket{\mathcal{S}_\mathfrak q,U}\\
   &= \int  r^2(\tau) d \tau d\mathbb{S}_2 \,    \mathcal T^{\mu \nu}_\phi(X(\tau)) \nonumber \,.
\end{align}
The tensor $\mathcal T^{\mu \nu}_\phi(X(\tau))$ encodes the annihilation of the qubit $\mathfrak q$ from the shell and the antiqubit $\bar{ \mathfrak q}$ from the $\ket{in}$ state. For concreteness, we consider the shell equipped with a qubit of a definite charge $-1/2$, so that $\bra{\mathcal S_{\cancel{\mathfrak{q}}}}\hat \phi(x) \ket{\mathcal S}= \phi_-(x)$. The conservation of the global $SU(2)$ charge during the scattering needs an \qu{antiqubit} with opposite charge to come from the $\ket{in}$ state $\ket{U}$, namely $\bra{U_{\cancel{\bar{\mathfrak{q}}}}} \hat \phi \ket{U}=\phi_+(x)$. In this way:
\begin{align}
    \mathcal T^{\mu \nu}_\phi(x)&= \nabla^\mu \phi_+ \nabla^\nu \phi_- + \nabla^\nu \phi_+ \nabla^\mu \phi_- \\
    &- \bar g^{\mu \nu} ( \nabla_\alpha \phi_+ \nabla^\alpha \phi_- - m^2 \phi_+ \phi_-) \,. \nonumber
\end{align}

Given the symmetry of the problem, the scalar field $\phi$ can be decomposed in spherical harmonics as is standard
\begin{equation}
    \phi(x)= \sum_{l,m} \frac{1}{r}Y_{lm}(\theta, \varphi) \chi_{lm}(r,v) \,,
\end{equation}
where $\chi_{lm}$ is a doublet in the fundamental representation of $SU(2)$. 

In this work we will focus solely on the $s$-wave component of the scattering ($l=m=0$). Additionally, we will retain the Wentzel--Kramers--Brillouin (WKB) approximation~\cite{Hall:2013jtz} for the radial part of the field, namely $Y_{00}(\theta, \varphi) = 1/\sqrt{4 \pi} $ and
\begin{equation}
    \chi_{+}(r,v)= \frac{e^{iS_+(r,v)}}{\sqrt{2 |\Omega_+|}} \,, \quad \chi_{-}(r,v)= \frac{e^{iS_-(r,v)}}{\sqrt{2 E_-}}\,,
\end{equation}
where $\chi_{00}=(\chi_+,\chi_-)$ and $E_-$ is the energy of the qubit in the local freely falling frame.

The point particle action $S_-$ for the qubit reads
\begin{align} 
    S_- (x)= -\int \omega dv + \int Q_r  dr \,,
\end{align}
where $\omega$ is the transferred energy between $in$ and $out$ shell-states and $Q_r$ is the correspondent transferred radial momentum. $E_-$ is the energy in the local frame $E_-= \dot X^\mu p_\mu= \omega \dot v + Q_r \dot r$ and the dot denotes the derivative with respect to the proper time $\tau$. 

Similarly, the action $S_+$ for the outgoing antiqubit is 
\begin{align} 
    S_+ (x)= \int \Omega_+ dv - \int k_r dr \,.
\end{align}
Near $F=0^-$, the outgoing qubit approximately moves on a null geodesic, with $k_r \simeq \Omega_+/F$.

\section{Scattering probability}

The process we want to describe is schematically the interaction of the qubit $\mathfrak q$ on the shell with the Unruh vacuum:
\begin{equation}
   \ket{U,\mathcal  S; M
   } \to \ket{U_{\cancel{\bar{\mathfrak{q}}}},\mathcal  S_{\cancel{\mathfrak{q}}}; M-\omega}\,.
\end{equation}
The matrix element \eqref{eq:Localized_T} implies an interaction through the stress energy tensor localized on the shell. In the $s$-wave approximation we get
\begin{align} \label{eq:amplitude}
    \mathcal{A}&= i\int d^4 x \sqrt{-g} \,\delta g_{\mu \nu}\bra{U_{\cancel{\bar{\mathfrak{q}}}},\mathcal S_{\cancel{\mathfrak{q}}}}  \hat T_\phi^{\mu \nu}\ket{U,\mathcal S
   } \nonumber\\
   & =i \int  r^2(\tau) d \tau d\mathbb{S}_2 \,   \delta g_{\mu \nu} \mathcal T^{\mu \nu}_\phi(X(\tau))
   \\
   &= i\int d \tau \left[ \frac{4 G\omega}{r} \nabla^v \chi_+ \nabla^v \chi_- \right]_{(r,v)=(r(\tau),v(\tau))} \,, \nonumber
\end{align}
where in the last step the integral over the 2-sphere $\int d\mathbb{S}_2=4 \pi$ has been performed. Note that
\begin{align}
      \nabla^v \chi_\pm = g^{\mu v} \nabla_\mu \chi_\pm= \partial_r\chi_\pm=i \chi_\pm(p_\pm)_r\,,
\end{align}
where $(p_\pm)_\mu:= \partial_\mu S_\pm$ is the derivative of the WKB phase. In particular, $(p_-)_r=Q_r$ is the momentum transferred to the shell and $(p_+)_r=k_r$ is the WKB momentum of the antiqubit. 

If the scattering happens in a patch of spacetime almost flat in the local inertial frame, we have that the rate per unit of proper time can be computed as usual \cite{Peskin:1995ev}:
\begin{align} \label{eq:rate}
     \Gamma(\tau)= \int d \Omega_+ n(\Omega_+) \int d \Pi_{\rm out}  \left| \frac{2 G\omega}{r(\tau)} \frac{k_r(\tau)Q_r }{\sqrt{|\Omega_+| E_-}}\right|^2 \\ \times \delta(Q_r-k_r(\tau))\delta(\omega+\Omega_+)\,, \nonumber
\end{align}
where $n(\Omega_+)d \Omega_+ $ is the number of antiqubits in the Unruh state in a shell $[\Omega_+,\, \Omega_+ +d \Omega_+]$ and $d\Pi_{\rm out}$ is the distribution of the final state of the shell, labeled by $Q_r$. The $\delta$-functions implementing the conservations of Killing energy and radial momentum can be understood in the following ways:
\begin{itemize}
    \item The conservation of the Killing energy amounts to requiring the two states $\ket{\mathcal{S},U}$ and $\ket{\mathcal{S}_{\cancel{\mathfrak{q}}},U_{\cancel{\mathfrak{\bar{q}}}}}$ to have the same ADM mass. In fact, the scattering process does not only modify the energy of the shell by \qu{subtracting} a quantum of energy $\omega$, but it also modifies the Unruh state so to generate an on-shell Hawking quanta outside of the trapping region, with energy $-\Omega_+$. The conservation of the ADM mass implies $\omega=-\Omega_+$.
    \item The conservation of the radial momentum is not due to a global statement. The system is not invariant under $r-$translation, so no notion of radial momentum is conserved. However, locally, the amplitude \eqref{eq:amplitude} maximizes in its saddle point, where the phase minimizes:
    \begin{align}
        &\frac{d}{d \tau} \left( Q_r r(\tau)- \int^{r(\tau)} k_r dr \right)=0 \\
        &\implies Q_r- k_r(r(\tau))=0 \nonumber .
    \end{align}
    That implies a local conservation rule along the radial direction.
\end{itemize}

It is easy to see that the rate in \cref{eq:rate} is enhanced for big values of $k_r$. Therefore, this process would be favoured in the near-horizon region, where $k_r \simeq \Omega_+/F$ diverges. Therefore, focusing in the region $|F| \ll 1$, we can also specify the occupation number of $in$ wavepackets, being \cite{Jacobson:2003vx}
\begin{align}
     n(\Omega_+)= \frac{1}{e^{|\Omega_+|/T_{\rm H}}-1} \,.
\end{align}
The density of $out$ states in terms of $Q_r$ can be written as
\begin{align}
     d \Pi_{\rm out}= \frac{dQ_r}{2 \pi((M-\omega) \dot v+ Q_r \dot r )} \,,
\end{align}
where the convention used for the normalization is $\braket{Q'_r|Q_r}=2 \pi 2E_{\mathcal S} \delta(Q_r-Q'_r)$, with $E_{\mathcal S}=(M-\omega) \dot v+ Q_r \dot r $ the energy of the shell in the local frame.

Plugging everything in, the near-horizon rate becomes
\begin{align} \label{eq:Gamma_NH}
   \Gamma(\tau) &= \frac{\omega^2}{M^2} \frac{1}{|\dot r|} \frac{n(\omega)}{2 \pi ((M-\omega) \dot v+ k_r(\tau) \dot r)} \frac{\omega^2}{F(r(\tau))^3}   \\
   &\simeq \frac{\omega^2}{M^2} \frac{n(\omega)}{ \dot r^2}  \frac{\omega}{F(r(\tau))^2} \nonumber\,,
\end{align}
where in the last step we considered the limit $k_r \simeq \Omega_+/F \gg M- \omega$.

\subsection{First scattering probability}
Having computed a rate per time unit, we are now able to compute the actual scattering probability. At first sight, one can recognize how the formula \eqref{eq:Gamma_NH} for the rate of scattering diverges for $F \to 0^-$. This mathematical feature is a consequence of the infinite blueshift experienced by the Unruh modes near the horizon \cite{Liberati:2009ak}. However, the physical quantity to be computed is the so-called \qu{first-scattering} probability. That is, the probability that the first process of this kind happens for some given $\tau>0$ (being $\tau=0$ the formation of the trapping region), without having happened at lower values of $\tau$. Once a single process happens, the background geometry changes and the computation must be iterated. The first-scattering probability is given by a Volterra type integral~\cite{Slavik2007}
\begin{align}
   P_1=1- \exp \left[-N \int_0^\tau \Gamma(\tau') d\tau'\right]\,,
\end{align}
where the factor $N$ accounts for having $N$ qubits of the same energy. 

The integral is dominated by the near horizon region for which
\begin{equation}
    F(r(\tau))=\frac{ \dot r(0)}{2GM} \tau+ \mathcal O(\tau^2) \,.
\end{equation}
Thus, the integrand becomes
\begin{align} \label{eq:rate_NH}
    \Gamma(\tau)\simeq4 G^2n(\omega) \frac{\omega^3}{\dot r^4}\biggl|_{\tau=0}  \frac{1}{\tau^2}\,.
\end{align}
Therefore,
\begin{equation}
    \int_0^\tau \Gamma(\tau') d\tau'=+\infty\,,
\end{equation}
and
\begin{equation}
    P_1=1\,,
\end{equation}
showing that this process must happen with unit probability.

On a final note, let us comment that the process is enhanced due to the high relative blueshift of the outgoing modes \cite{Liberati:2009ak}. One might be worried about the trans-Planckian nature of such modes. The presence of these, however, is an essential feature in the standard derivation of \cite{Hawking:1975vcx}, and so trans-Planckian modes are required to formulate the information-loss problem. Hence, questioning this result, on the basis of the trans-Planckian problem, amounts to questioning Hawking evaporation and the information-loss problem to start with. 

For instance, the implementetion of a hard UV cutoff directly spoils the computation performed in \cite{Hawking:1975vcx}. It would nevertheless be important to investigate this mechanism in scenarios where an appropriate implementation of the cutoff prevents modes from reaching trans-Planckian frequencies without spoiling the Hawking evaporation process, as in \cite{Brout:1995wp,DelPorro:2023lbv,DelPorro:2026grx}.

\section{Physical interpretation}

The process analysed in the previous section is a local process happening in the interior of the trapped region. 
Indeed, as soon as the horizon forms, the system has access to negative-normed modes. The divergent nature of the probability rate \eqref{eq:rate} implies that the scattering arises arbitrarily close the horizon. Furthermore, the scattering carries out a finite amount of energy $\omega$, thus the trapped region locally disappears. 

The non-local nature of the quantum state implies that this process affects the exterior geometry. 
First of all, it is straightforward to show that the annihilation of an Hawking partner inside the horizon with energy $-\omega$ accounts for the creation of an Hawking quanta outside the horizon with energy $\omega$. In fact, in the Unruh state \eqref{eq:U_state} the following relation holds \cite{Jacobson:2003vx},
\begin{equation}
    a_{-\omega,\pm}^{\rm in}\ket{U}=e^{-\pi\omega/\kappa}a_{\omega,\mp}^{\dagger,{\rm out}}\ket{U}\,.
\end{equation}
Note that the $SU(2)$ charge of the $out$ mode has opposite sign of the one of the $in$ mode and thus it is the same of the qubit $\mathfrak{q}$ destroyed on the shell. Thus, the process transfers the whole information of the qubit from the collapsing shell to an outgoing mode outside the horizon\footnote{Due to the linearity of the process, if the qubit is in a linear superposition of positive and negative charge, the outgoing mode would be in the same linear superposition. }.  

\begin{figure}
    \centering
    \includegraphics[width=0.48\linewidth]{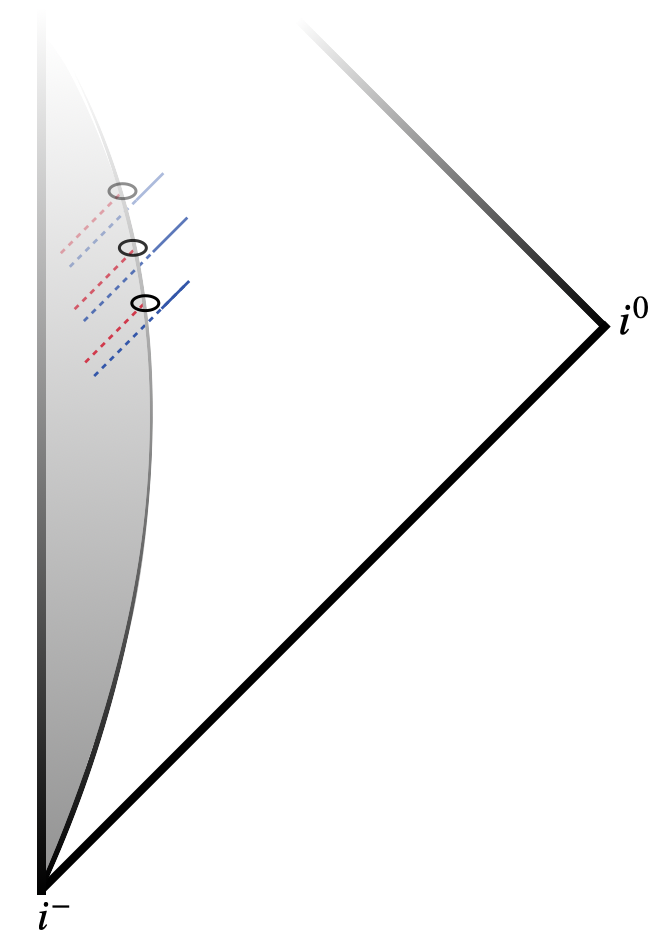}
    \caption{Causal diagram considering the backreaction of the interaction. The virtual particle (red dashed lines) interact with a qubit on the shell and closes the trapping horizon (black circles). The energy is transmitted to the outgoing Hawking quanta (blue lines). The endpoint is not shown as it is beyond the capability of the current approximation.}
    \label{fig:With_backreaction}
\end{figure}

After a single scattering, the probability of the Hawking quanta to escape at infinity is governed, as usual, by the gray-body factor~\cite{Fabbri:2005mw}.
At the same time, the (lightened) shell keeps falling freely and will eventually form again a trapped region. This gives new access to negative-normed modes, and the process can kick in again, as depicted in Fig.~\ref{fig:With_backreaction}. We note that our treatment can iteratively be applied as long as the following approximations hold:
\begin{enumerate}
    \item The linearized term of \cref{eq:interaction_L} contributes as the leading order to the process. This is true when $\delta g^2 \ll \delta g$ or, equivalently, when $\omega \ll M$. 
    \item The vacuum state is given by the Unruh state. After each process, the collapsing matter configuration and the quantum state change as the consequence of the annihilation processes. We point out, however, that the details of the matter quantum state are not essential, provided that a reservoir $n(\omega)$ of negative-normed modes can be accessed by the shell.
\end{enumerate}
Both conditions are approximately satisfied as long as the fraction of extracted energy is small compared to the total energy of the shell.
We do not expect the process to stop once these approximations are violated. Determining its details is beyond the scope of this work.
However, we can indulge in speculation about the possible final state of the collapse. As we described it, the shell-vacuum interaction we provided acts as a gravitational quantum friction for the freely falling matter, which stands against the formation of the trapping horizon. As a consequence, we see at least three possible scenarios
\begin{enumerate}
    \item Complete evaporation: The process might continue indefinitely. The emitted energy evaporates towards infinity.
    \item Horizonless ultracompact object: The process effeciently prevents the formation of a trapped region; (part of) the radiated energy stays close to the would-be horizon. The endpoint of the dynamics would be an horizonless object supported by semiclassical effects \cite{Carballo-Rubio:2017tlh,Arrechea:2021xkp,Arrechea2024}. 
    \item Formation of singularity: The process switches off once the approximations listed above stop being satisfied. The collapse then proceeds until a singularity is formed.
\end{enumerate}
Although far from being definitive, it is intriguing that the possibility of a singularity-free object may arise naturally from minimal semiclassical assumptions, with no new physics involved in the set up. Note that the possibility of having an horizonless gravitational bound state rather than a completely evaporated object might depend on the initial condition of the collapse. We emphasize the difference with respect to the semiclassical gravitational collapse studied in the literature \cite{Parentani:1994ij,Barcelo:2007yk}, where is it implicitly assumed that collapsing matter cannot interact with the vacuum state.

\section{Discussion}
In this Letter we have analysed the gravitational collapse in general relativity with minimal semiclassical assumptions.
We considered a simple toy model of a thin shell made of scalar particles $\phi$ charged under a global $SU(2)$ symmetry. 

We showed that the presence of trans-Plankian modes, which leads to Hawking evaporation and the information-loss problem, also generates a \textit{gravitational quantum friction} opposing the formation of a trapping surface. 

This qualitatively new phenomenon can have crucial implications for several open problems.
To start, we show that the non-local nature of the vacuum state allows to transfer information outside the trapped region, with direct implications on the information-loss problem. 

Furthermore, the use of a global symmetry to model the information of the system, also shows that the evaporation can conserve global charges questioning the assumptions of the no-global symmetries conjectures~\cite{Kallosh:1995hi,Harlow:2018tng,Harlow:2020bee}. 

Although different and more realistic matter contents and collapsing geometries could be considered, the picture emerging from our simple example is solid and well-controlled by our approximations.

Finally, we discussed a speculative end-point of the collapse, arguing that a better understanding of semiclassical effects can have important repercussions on the formation of the singularity, even before any full quantum gravity theory is needed \cite{DiFilippo:2025kzh}.

Remarkably, without any ingredient beyond the standard semiclassical picture, we showed that the interaction between the collapsing matter and the modes in the vacuum state cannot be ignored to have a self-consistent dynamical picture. The inclusion of such interaction can change our understanding of some of the most compelling open problems in gravitational physics.

\section*{Acknowledgments}
We thank Luca Buoninfante, Raúl Carballo-Rubio, Stefano Liberati, and Eduardo Martín-Martínez and Alessia Platania for insightful discussions.
FDP acknowledges support of the research grant (VIL60819) from VILLUM FONDEN. The Center
of Gravity is a Center of Excellence funded by the Danish National Research Foundation under grant No. 184. FDF acknowledges financial support from the European Union’s Horizon 2020 research and innovation programme under the Marie Sklodowska-Curie Actions (grant agreement ID:101207584)

\bibliography{refs}

@inproceedings{Jacobson:2003vx,
    author = "Jacobson, Ted",
    title = "{Introduction to quantum fields in curved space-time and the Hawking effect}",
    booktitle = "{School on Quantum Gravity}",
    eprint = "gr-qc/0308048",
    archivePrefix = "arXiv",
    doi = "10.1007/0-387-24992-3_2",
    pages = "39--89",
    month = "8",
    year = "2003"
}

@article{Hawking:1975vcx,
    author = "Hawking, S. W.",
    editor = "Gibbons, G. W. and Hawking, S. W.",
    title = "{Particle Creation by Black Holes}",
    doi = "10.1007/BF02345020",
    journal = "Commun. Math. Phys.",
    volume = "43",
    pages = "199--220",
    year = "1975",
    note = "[Erratum: Commun.Math.Phys. 46, 206 (1976)]"
}

@article{Liberati:2009ak,
    author = "Liberati, Stefano and Sindoni, Lorenzo and Sonego, Sebastiano",
    title = "{Linking the trans-Planckian and the information loss problems in black hole physics}",
    eprint = "0904.0815",
    archivePrefix = "arXiv",
    primaryClass = "gr-qc",
    doi = "10.1007/s10714-009-0899-2",
    journal = "Gen. Rel. Grav.",
    volume = "42",
    pages = "1139--1152",
    year = "2010"
}

@article{Agullo:2024nxg,
    author = "Agullo, Ivan and Cabrera, Paula Calizaya and Elizaga Navascu{\'e}s, Beatriz",
    title = "{Entangled pairs in evaporating black holes without event horizons}",
    eprint = "2407.03031",
    archivePrefix = "arXiv",
    primaryClass = "gr-qc",
    doi = "10.1103/PhysRevD.110.085002",
    journal = "Phys. Rev. D",
    volume = "110",
    number = "8",
    pages = "085002",
    year = "2024"
}

@article{Susskind:1993if,
    author = "Susskind, Leonard and Thorlacius, Larus and Uglum, John",
    title = "{The Stretched horizon and black hole complementarity}",
    eprint = "hep-th/9306069",
    archivePrefix = "arXiv",
    reportNumber = "SU-ITP-93-15",
    doi = "10.1103/PhysRevD.48.3743",
    journal = "Phys. Rev. D",
    volume = "48",
    pages = "3743--3761",
    year = "1993"
}

@article{Almheiri:2012rt,
    author = "Almheiri, Ahmed and Marolf, Donald and Polchinski, Joseph and Sully, James",
    title = "{Black Holes: Complementarity or Firewalls?}",
    eprint = "1207.3123",
    archivePrefix = "arXiv",
    primaryClass = "hep-th",
    doi = "10.1007/JHEP02(2013)062",
    journal = "JHEP",
    volume = "02",
    pages = "062",
    year = "2013"
}

@article{Almheiri:2019yqk,
    author = "Almheiri, Ahmed and Mahajan, Raghu and Maldacena, Juan",
    title = "{Islands outside the horizon}",
    eprint = "1910.11077",
    archivePrefix = "arXiv",
    primaryClass = "hep-th",
    month = "10",
    doi = "",
    journal = "",
    volume = "",
    pages = "",
    year = "2019"
}

@article{Almheiri:2019psf,
    author = "Almheiri, Ahmed and Engelhardt, Netta and Marolf, Donald and Maxfield, Henry",
    title = "{The entropy of bulk quantum fields and the entanglement wedge of an evaporating black hole}",
    eprint = "1905.08762",
    archivePrefix = "arXiv",
    primaryClass = "hep-th",
    doi = "10.1007/JHEP12(2019)063",
    journal = "JHEP",
    volume = "12",
    pages = "063",
    year = "2019"
}

@article{Mathur:2005zp,
    author = "Mathur, Samir D.",
    editor = "Kiritsis, E.",
    title = "{The Fuzzball proposal for black holes: An Elementary review}",
    eprint = "hep-th/0502050",
    archivePrefix = "arXiv",
    doi = "10.1002/prop.200410203",
    journal = "Fortsch. Phys.",
    volume = "53",
    pages = "793--827",
    year = "2005"
}

@article{Giddings:1993vj,
    author = "Giddings, Steven B.",
    title = "{Comments on information loss and remnants}",
    eprint = "hep-th/9310101",
    archivePrefix = "arXiv",
    reportNumber = "UCSBTH-93-35",
    doi = "10.1103/PhysRevD.49.4078",
    journal = "Phys. Rev. D",
    volume = "49",
    pages = "4078--4088",
    year = "1994"
}

@article{Unruh:2017uaw,
    author = "Unruh, William G. and Wald, Robert M.",
    title = "{Information Loss}",
    eprint = "1703.02140",
    archivePrefix = "arXiv",
    primaryClass = "hep-th",
    doi = "10.1088/1361-6633/aa778e",
    journal = "Rept. Prog. Phys.",
    volume = "80",
    number = "9",
    pages = "092002",
    year = "2017"
}

@inproceedings{Buoninfante:2025gqk,
    author = "Buoninfante, Luca and Di Filippo, Francesco",
    title = "{Is the information loss problem a paradox?}",
    eprint = "2504.00516",
    archivePrefix = "arXiv",
    primaryClass = "gr-qc",
    month = "4",
    year = "2025"
}

@article{Buoninfante:2021ijy,
    author = "Buoninfante, Luca and Di Filippo, Francesco and Mukohyama, Shinji",
    title = "{On the assumptions leading to the information loss paradox}",
    eprint = "2107.05662",
    archivePrefix = "arXiv",
    primaryClass = "hep-th",
    reportNumber = "YITP-21-76, IPMU21-0049",
    doi = "10.1007/JHEP10(2021)081",
    journal = "JHEP",
    volume = "10",
    pages = "081",
    year = "2021"
}

@article{Hawking:1993pb,
    author = "Hawking, S. W.",
    title = "{The Superscattering matrix for two-dimensional black holes}",
    eprint = "hep-th/9401109",
    archivePrefix = "arXiv",
    doi = "10.1103/PhysRevD.50.3982",
    journal = "Phys. Rev. D",
    volume = "50",
    pages = "3982--3986",
    year = "1994"
}

@article{Mathur:2009hf,
    author = "Mathur, Samir D.",
    editor = "Uranga, A. M.",
    title = "{The Information paradox: A Pedagogical introduction}",
    eprint = "0909.1038",
    archivePrefix = "arXiv",
    primaryClass = "hep-th",
    doi = "10.1088/0264-9381/26/22/224001",
    journal = "Class. Quant. Grav.",
    volume = "26",
    pages = "224001",
    year = "2009"
}

@book{Wald:1984rg,
    author = "Wald, Robert M.",
    title = "{General Relativity}",
    doi = "10.7208/chicago/9780226870373.001.0001",
    publisher = "Chicago Univ. Pr.",
    address = "Chicago, USA",
    year = "1984"
}

@Inbook{Arrechea2024,
author="Arrechea, Julio
and Barcel{\'o}, Carlos
and Boyanov, Valentin",
editor="Malafarina, Daniele
and Joshi, Pankaj S.",
title="After Collapse: On How a Physical Vacuum Can Change the Black Hole Paradigm",
bookTitle="New Frontiers in Gravitational Collapse and Spacetime Singularities",
year="2024",
publisher="Springer Nature Singapore",
address="Singapore",
pages="1--51",
isbn="978-981-97-1172-7",
doi="10.1007/978-981-97-1172-7_1",
url="https://doi.org/10.1007/978-981-97-1172-7_1"
}

@article{Carballo-Rubio:2019fnb,
    author = "Carballo-Rubio, Ra{\'u}l and Di Filippo, Francesco and Liberati, Stefano and Visser, Matt",
    title = "{Geodesically complete black holes}",
    eprint = "1911.11200",
    archivePrefix = "arXiv",
    primaryClass = "gr-qc",
    doi = "10.1103/PhysRevD.101.084047",
    journal = "Phys. Rev. D",
    volume = "101",
    pages = "084047",
    year = "2020"
}

@article{Carballo-Rubio:2025fnc,
    author = "Carballo-Rubio, Ra{\'u}l and others",
    title = "{Towards a non-singular paradigm of black hole physics}",
    eprint = "2501.05505",
    archivePrefix = "arXiv",
    primaryClass = "gr-qc",
    doi = "10.1088/1475-7516/2025/05/003",
    journal = "JCAP",
    volume = "05",
    pages = "003",
    year = "2025"
}

@article{Arrechea:2021xkp,
    author = "Arrechea, Julio and Barcel{\'o}, Carlos and Carballo-Rubio, Ra{\'u}l and Garay, Luis J.",
    title = "{Semiclassical relativistic stars}",
    eprint = "2110.15808",
    archivePrefix = "arXiv",
    primaryClass = "gr-qc",
    doi = "10.1038/s41598-022-19836-8",
    journal = "Sci. Rep.",
    volume = "12",
    number = "1",
    pages = "15958",
    year = "2022"
}

@article{Carballo-Rubio:2017tlh,
    author = "Carballo-Rubio, Ra{\'u}l",
    title = "{Stellar equilibrium in semiclassical gravity}",
    eprint = "1706.05379",
    archivePrefix = "arXiv",
    primaryClass = "gr-qc",
    doi = "10.1103/PhysRevLett.120.061102",
    journal = "Phys. Rev. Lett.",
    volume = "120",
    number = "6",
    pages = "061102",
    year = "2018"
}

@article{DiFilippo:2025kzh,
    author = "Di Filippo, Francesco",
    title = "{Radiating Black Holes in General Relativity Need Not Be Singular}",
    eprint = "2510.20649",
    archivePrefix = "arXiv",
    primaryClass = "gr-qc",
    doi = "10.1103/gv8z-f128",
    journal = "Phys. Rev. Lett.",
    volume = "136",
    number = "16",
    pages = "161402",
    year = "2026"
}

@article{Barcelo:2007yk,
    author = "Barcelo, Carlos and Liberati, Stefano and Sonego, Sebastiano and Visser, Matt",
    title = "{Fate of gravitational collapse in semiclassical gravity}",
    eprint = "0712.1130",
    archivePrefix = "arXiv",
    primaryClass = "gr-qc",
    doi = "10.1103/PhysRevD.77.044032",
    journal = "Phys. Rev. D",
    volume = "77",
    pages = "044032",
    year = "2008"
}

@book{Fabbri:2005mw,
    author = "Fabbri, A. and Navarro-Salas, J.",
    title = "{Modeling black hole evaporation}",
    doi = "10.1142/p378",
    isbn = "978-1-86094-527-4, 978-1-86094-722-3, 978-1-78326-038-6",
    publisher = "World Scientific",
    address = "Singapore",
    year = "2005"
}

@article{Page:1993wv,
    author = "Page, Don N.",
    title = "{Information in black hole radiation}",
    eprint = "hep-th/9306083",
    archivePrefix = "arXiv",
    reportNumber = "ALBERTA-THY-24-93",
    doi = "10.1103/PhysRevLett.71.3743",
    journal = "Phys. Rev. Lett.",
    volume = "71",
    pages = "3743--3746",
    year = "1993"
}

@article{Wald:1975kc,
    author = "Wald, Robert M.",
    title = "{On Particle Creation by Black Holes}",
    doi = "10.1007/BF01609863",
    journal = "Commun. Math. Phys.",
    volume = "45",
    pages = "9--34",
    year = "1975"
}

@book{Hall:2013jtz,
    author = "Hall, Brian C.",
    title = "{Quantum Theory for Mathematicians}",
    doi = "10.1007/978-1-4614-7116-5",
    year = "2013"
}

@book{Peskin:1995ev,
    author = "Peskin, Michael E. and Schroeder, Daniel V.",
    title = "{An Introduction to quantum field theory}",
    doi = "10.1201/9780429503559",
    isbn = "978-0-201-50397-5, 978-0-429-50355-9, 978-0-429-49417-8",
    publisher = "Addison-Wesley",
    address = "Reading, USA",
    year = "1995"
}

@book{Slavik2007,
author = {Slavík, Antonín},
location = {Praha},
publisher = {Matfyzpress},
title = {Product integration. Its history and applications},
url = {http://eudml.org/doc/202418},
year = {2007},
}

@article{Brout:1995wp,
    author = "Brout, R. and Massar, S. and Parentani, R. and Spindel, P.",
    title = "{Hawking radiation without transPlanckian frequencies}",
    eprint = "hep-th/9506121",
    archivePrefix = "arXiv",
    reportNumber = "ULB-TH-95-06, UMH-MG-95-02, LPTENS-95-18",
    doi = "10.1103/PhysRevD.52.4559",
    journal = "Phys. Rev. D",
    volume = "52",
    pages = "4559--4568",
    year = "1995"
}

@article{DelPorro:2023lbv,
    author = "Del Porro, F. and Herrero-Valea, M. and Liberati, S. and Schneider, M.",
    title = "{Hawking radiation in Lorentz violating gravity: a tale of two horizons}",
    eprint = "2310.01472",
    archivePrefix = "arXiv",
    primaryClass = "gr-qc",
    doi = "10.1007/JHEP12(2023)094",
    journal = "JHEP",
    volume = "12",
    pages = "094",
    year = "2023"
}

@article{DelPorro:2026grx,
    author = "Del Porro, Francesco and Liberati, Stefano and Schneider, Marc",
    title = "{Hawking radiation with dispersion: reconciling the Bogoliubov and tunneling approaches}",
    eprint = "2604.24861",
    archivePrefix = "arXiv",
    primaryClass = "gr-qc",
    month = "4",
    journal="",
    year = "2026"
}

@article{Kallosh:1995hi,
    author = "Kallosh, Renata and Linde, Andrei D. and Linde, Dmitri A. and Susskind, Leonard",
    title = "{Gravity and global symmetries}",
    eprint = "hep-th/9502069",
    archivePrefix = "arXiv",
    reportNumber = "SU-ITP-95-2",
    doi = "10.1103/PhysRevD.52.912",
    journal = "Phys. Rev. D",
    volume = "52",
    pages = "912--935",
    year = "1995"
}

@article{Harlow:2018tng,
    author = "Harlow, Daniel and Ooguri, Hirosi",
    title = "{Symmetries in quantum field theory and quantum gravity}",
    eprint = "1810.05338",
    archivePrefix = "arXiv",
    primaryClass = "hep-th",
    doi = "10.1007/s00220-021-04040-y",
    journal = "Commun. Math. Phys.",
    volume = "383",
    number = "3",
    pages = "1669--1804",
    year = "2021"
}

@article{Harlow:2020bee,
    author = "Harlow, Daniel and Shaghoulian, Edgar",
    title = "{Global symmetry, Euclidean gravity, and the black hole information problem}",
    eprint = "2010.10539",
    archivePrefix = "arXiv",
    primaryClass = "hep-th",
    doi = "10.1007/JHEP04(2021)175",
    journal = "JHEP",
    volume = "04",
    pages = "175",
    year = "2021"
}

@book{Birrell:1982ix,
    author = "Birrell, N. D. and Davies, P. C. W.",
    title = "{Quantum Fields in Curved Space}",
    doi = "10.1017/CBO9780511622632",
    isbn = "978-0-511-62263-2, 978-0-521-27858-4",
    publisher = "Cambridge University Press",
    address = "Cambridge, UK",
    series = "Cambridge Monographs on Mathematical Physics",
    year = "1982"
}

@article{Unruh:1976db,
    author = "Unruh, W. G.",
    title = "{Notes on black hole evaporation}",
    doi = "10.1103/PhysRevD.14.870",
    journal = "Phys. Rev. D",
    volume = "14",
    pages = "870",
    year = "1976"
}

@article{Parentani:1994ij,
    author = "Parentani, Renaud and Piran, Tsvi",
    title = "{The Internal geometry of an evaporating black hole}",
    eprint = "hep-th/9405007",
    archivePrefix = "arXiv",
    doi = "10.1103/PhysRevLett.73.2805",
    journal = "Phys. Rev. Lett.",
    volume = "73",
    pages = "2805--2808",
    year = "1994"
}
\end{document}